# An Application of Chaos Theory for Estimation of Simultaneous Variability of RR-intervals in Heart and Systolic Blood Pressure in Humans

Elio Conte[1,2], Antonio Federici [1], Joseph P. Zbilut[3]

[1]*Department of Pharmacology and Human Physiology and Tires, Center for Innovative Technologies for Signal Detection and Processing, University of Bari, Italy*
[2]*School of Advanced International Studies on Theoretical and nonLinear Methodologies in Physics-Bari, Italy*
[3]*Department of Molecular Biophysics and Physiology, Rush University Medical Center, 1653W Congress, Chicago, IL60612, USA*.

**Abstract:** we introduce a new method to estimate BaroReflex Sensitivity (BRS). The methodology, based on the CZF formulation, recently published (see Conte et al 2008), enables to evaluate simultaneous variability of RR and SBP and to estimate the coupling strength. The technique is applied to subjects (female and men with age ranging from 21 to 28 years old) and it is compared with the results that may be obtained by using the standard Fourier spectral analysis technique. The comparison is also performed by using the technique of Lomb-Scargle periodogram, based on Fourier analysis.

## 1. On the Fundamental Role of Recurrences and Variability of Signals in Nature

Only few systems in Nature exhibit linearity. The greatest set of natural systems, especially those who pertain to biological matter, to physiological, neuro-physiological and to psychological processes, possess a complexity that results in a great variedness and variability, linked to non linearity, to non stationarity, and to non predictability of their time dynamics. In time domain, traditional methods were first used to describe the amplitude distribution of signals. Later, methodologies used spectral analysis. Unfortunately, they suffer of fundamental limits. They are applied assuming linearity and stationarity of signals that actually do not exist. The consequence is that such methods are unable to analyse in a proper way the irregularity present in most of signals. Really, such irregularity is the basis of the dynamics that we explore. It reveals that it has complex behaviours. They are very distant from traditionally accepted principles as it is the case of homeostatic equilibrium and similar mechanisms of controls. The study of this very irregular behaviour requires the introduction of new basic principles. Non linear science is becoming an emerging methodological and theoretical framework that makes up what is called the science of the complexity, often called also chaos theory. The aim of non linear methodologies is a description of complexity and the exploration of the multidimensional interactions within and among components of given systems. An important concept here is that one of chaotic or divergent behaviour. It will be defined chaotic if trajectories issuing from points of whatever degree of proximity in the space of phase, distance themselves from one another over time in an exponential way.
In detail, the basic principles may be reassumed as it follows:
1) Non linear systems under certain conditions may exhibit chaotic behaviour.
2) The behaviour of a chaotic system can change drastically in response to small changes in the system's initial conditions;
3) A chaotic system is often deterministic;
4) In chaotic systems the output system is no more proportionate to system input.

Chaos may be identified in systems also excluding the requirement of determinism. We term such behaviour non-deterministic chaos. This approach to chaos theory was initiated by Zak, Zbilut and Webber [Zak 1989,1992,1998; Zbilut et al. 1994, 1995, 1996, Zbilut et al. 2008] and recently we have given several examples, theoretical and experimental verifications on this important chaotic behaviour [Conte et al. 2004, 2004, 2006].

## 2. Fractality and Non Linearity of Experimental Time Series

Complexity of natural processes relates the variedness and the variability of the experimentally measured signals in the form of time series. The CZF method that we introduced rather recently [Conte et al. 2007, 2008, 2008, Mastrolonardo et al. 2006] for analysis and quantification of sympathetic and vagal activity in variability of R–R time series in ECG signals, relates such feature. It states for the surname (Conte, Zbilut, Federici) of the authors who introduced it. Let us explain the essential reasons that motivate the introduction of CZF method.

Let us recall an old notion. The presence of an harmonic component in a given time series is revealed by its power spectrum $P(v)$ given by the squared norm of the Fourier transform of the given time series $X(t)$ as

$$P(v) = \left\| \int_0^\infty e^{ivt} X(t) dt \right\|^2 \qquad (2.1)$$

and evidencing a single sharp peak.

FFT (Fast Fourier Transform), in its discrete version, is currently applied in analysis of non linear time series. Because of its simplicity, Fourier analysis has dominated and still dominates the data analysis efforts. This happens ignoring that FFT is valid under extremely general conditions but essentially under the respect of some crucial restrictions that often result largely violated, especially for electrophysiological signals. Three stringent conditions must be observed for FFT:
1) the system must be linear.
   Instead the data that we analyse are non linear time series of signals.
2) The data of the time series must be strictly periodic and stationary.
   Instead, the feature of signals under our investigation, is that they show irregular dynamics, and rarely exhibit stationarity.
3) All the data of the time series must be sampled at equally spaced time intervals.
   Also this occurrence happens rarely, particularly for electrophysiological signals.

The consequences of such improper use of FFT are significant. The presence of non linearity and of non stationarity gives little sense to the results that are obtained by FFT.

We may now reassume some feature of the non linear method, that we have called CZF exposed in detail elsewhere [Conte et al. 2007, 2008, 2008, Mastrolonardo et al. 2006].

The study of stochastic processes with power-law spectra started with the paper on fractional Brownian motion (fBm) by Mandelbrot and Van Ness in 1968 [Mandelbrot et al. 1968, 1982] Fixed the initial conditions, fBm is defined by the following equation

$$X(ht) \stackrel{d}{=} h^H X(t) \qquad (2.2)$$

Given a self-similar fractal time series, the (2.1) establishes that the distribution remains unchanged by the factor $h^H$ even after the time scale is changed. ($\stackrel{d}{=}$) states that the statistical distribution function remains unchanged. $H$ is called Hurst exponent [Hurst et al.1965]. It varies as $0 < H < 1$, and it characterizes the general power – law scaling. For an additive process of Gaussian white noise, we have $H = 0.5$. $H$ values greater than 0.5 indicate persistence in time series (past trend persist into the future(long-range correlation)). $H$ values less than 0.5 indicate antipersistence (past trends tend to reverse in the future). The fBm also exhibits power-law behaviour in the Fourier spectrum. There is a linear relationship between the log of spectral power vs. log of frequency. The inverse of the slope in the log-log plot is called the spectral exponent $\beta$ ($1/f^\beta$ behaviour), and it is related to $H$ by the following relationship

$$H = \frac{\beta - 1}{2}$$

Let us start with Hurst analysis [Hurst et al.1965].] that brings light on some statistical properties of time series $X(t)$ that scales with an observed period of observation $T$ and a time resolution $\mu$. Scaling results characterized by an exponent $H$ that relates the long-term statistical dependence of the signal. In substance, one may generalize such Hurst approach, expressing the scaling behaviour of statistically significant properties of the signal. Indicating by $E$ the mean values, we have to analyze the q-order moments of the distribution of the increments

$$K_q(\tau) = \frac{E(|X(t+\tau) - X(t)|^q)}{E(|X(t)|^q)} \quad (2.3)$$

The (2.3) represents the statistical time evolution of the given stochastic variable $X(t)$.
For q=2, we may re-write the (2.3) in the following manner

$$\gamma(h) = \frac{1}{2n(h)} \sum_{i=1}^{n(h)} [X(u_{i+h}) - X(u_i)]^2 \quad (2.4)$$

that estimates the variogram of the given time series. Here, $n(h)$ is the number of pairs at lag distance $h$ while $X(u_i)$ and $X(u_{i+h})$ are time sampled series values at times $t$ and $t+h$, $t = u_1, u_2, ....; h = 1, 2, 3, .....$. In substance, the variogram is a statistical measure expressed in the form:

$$\gamma(h) = \frac{1}{2} Var[X(u+h) - X(u)] \quad (2.5)$$

The Variogram, here introduced, represents the most valuable measure of complexity of a given non linear time series and at the same time it enables us to overcome the difficulties previously mentioned for use of FFT. The concept of variability is sovereign in this case. Let us make an example to illustrate its relevance.
Let us admit we have a time series given only by six terms:
$X_1, X_2, X_3, X_4, X_5, X_6$. $\quad (2.6)$
The first time we select a time lag $h = 1$, and, using the (2.4), we calculate variability of this signal at this time scale, obtaining:
$(X_1 - X_2)^2 + (X_2 - X_3)^2 + (X_3 - X_4)^2 + (X_4 - X_5)^2 + (X_5 - X_6)^2 \quad (2.7)$
This is the variability of the signal at time scale $h = 1$ and, in accord with the (2.4), we indicate it by $\gamma_1(h) \equiv \gamma_1(1)$.
Note some important features:
The differences $(X_i - X_{i+1})^2$ in the (2.7) will account directly of the fluctuations (and thus of the total variability) that intervene in $X_{i+1}$ respect to $X_i$. It will be due to the particular features of the dynamics under investigation. Let us consider as example the case of the (2.6) representing the beat-to-beat fluctuations of human heartbeat intervals. The (2.7) will represent the total variability in time lag $h = 1$ due to the regulative activity exercised from sympathetic, vagal, and VLF activities in the time lag considered.
If $\gamma_1(1)$ will assume a value going to zero, we will conclude that at such time scale (time lag delay h = 1) the variability of the signal in this time lag is very modest. Otherwise, if $\gamma_1(1)$ will result different from zero in a consistent way, we will conclude that it gives great variability, to be attributed to the presence of a relevant activity of control. Still, the count of such variability will happen for all the points of the given time series and thus it will account for the total variability at the fixed time scale of resolution for the whole considered R–R process.

After to have computed the total variability of signal at this time resolution $h=1$, we will continue our calculation evaluating this time the total variability of the signal at the time resolution $h=2$, and thus calculating $\gamma_2(2)$. In the similar way we will proceed calculating total variability at the time scale resolution corresponding to $h=3$ and so on, completing the analysis of variability at each time scale. In conclusion we calculate the total variability of the signal, step by step, at different time scales. The result will be a diagram in a plot in which in axis of the ordinate we have the values of variability while in the axis of the abscissa we have the corresponding value of $h$, that is to say of the corresponding time resolution.

Note that in the standard cases we have excellent software to perform a complete HRV analysis. This is the case of HRV Analysis software [Juka-Pekka Niskanen et al. 2004].

To complete our summary on CZF method we still outline that, in calculating the $\gamma_i(h)$, we may also use the embedding procedure for reconstruction in phase space of the given time series.

By the CZF method we may also perform fractal analysis. In fact we may use the Fractal Variance Function, $\gamma(h)$, and the Generalized Fractal Dimension, $D_{dim}$, by the following equation

$$\gamma(h) = Ch^{D_{dim}} \quad (2.8)$$

and estimating the Marginal Density Function for self-affine distributions, given by the following equation [Wei et al. 2005, Conte et al. 2007, 2008, 2008, Mastrolonardo et al. 2006]:

$$P(h) = ak^{-a}h^{a-1} \quad (2.9)$$

This last consideration completes our brief view on CZF method. In detail, we will apply it

R-R time series that are related the beat-to-beat time fluctuations of human heartbeat intervals. R-R values are largely controlled by various physiological factors and, in particular, by the balance between sympathetic and parasympathetic nervous system activity imposed upon the spontaneous discharge frequency of the sinoatrial node. Fluctuations in time in R-R give origin to what we call the variability of the R-R signal and, using FFT, in frequency domain three bands are identified. The first, the VLF, is usually considered to range from 0 to 0.04 Hz and is related to humoral regulation of the sinus pacemaker cell activity; the second, the LF, ranging from 0.04 to 0.15 Hz, and the HF, ranging from 0.15 to 0.4 Hz are related to autonomic sympathetic and vagal activities, respectively.

To perform analysis of variability by CZF in frequency domain we calculate the mean value, $E(R-R)$, in msec.

We estimate an equivalent frequency

$$f_{equivalent} = \frac{1}{E(R-R)} \quad (2.10)$$

and we realize a diagram having in ordinate the values of the variability as calculated by the (2.4) and in abscissa, in correspondence of each lag, $h$, we will sign instead the value $hf_{equivalent}$

with $h=1,2,3,....$

### 3. Estimation of Baro Reflex Sensitivity (BRS) by the CZF Method.

The baroreflex loop is an important cardiovascular control mechanism for short-term blood pressure (BP) regulation. Based on afferent information of arterial baroreceptors reacting on changes in BP, central cardiovascular control is exerted on different peripheral effector systems as in particular on heart rate, cardiac output, peripheral resistance in order to keep BP between narrow limits. Baroreflex sensitivity (BRS) is a sensitive integrated measure of both sympathetic and parasympathetic autonomic regulation in which changes in heart rate due to variation in BP are reflected.

Different techniques, based usually on spectral analysis or on the so called sequence method, have been introduced to quantify baroreflex gain (Parlow et al. 1995). Traditionally, BRS is assessed pharmacologically, using the heart rate response to vasoactive drugs. Pharmacological and non-invasive BRS measurements have been found to correlate significantly (Parlow et al. 1995, Watkins et al. 1996). The literature is boundless on this subject. However, no definitive agreement has been reached on which of the employed methods should be preferred (Lipman et al. 2003, Parati et al. 2004). On the other hand, BRS measurements represent an important prognostic tool to detect early subclinical autonomic dysfunction. In fact, reduced values of BRS may result largely from vagal withdrawal determining an high component of risk. Therefore it represents a valuable predictor of future cardiovascular morbidity and mortality in a variety of disease states that are associated with autonomic failure, (Gerritsen et al. 2001, La Rovere et al. 1998, Ditto 1990). An interesting connection has been shown between diminished BRS and psychopathology (Virtanen et al. 2003). We retain that one of the reasons on the existing uncertainty in estimating BRS resides in the fact that all the employed method use linear approaches where the problem has instead an intrinsic non linear and non stationary dynamics. This is the reason for which we suggest here the introduction of a new method that accounts for variability, and essentially for non linearity and non stationarity of the investigated data. The method we propose is essentially free from approximations. Therefore, also if it is presently applied only to a preliminary investigation, it should be considered of valuable importance in order to execute a correct estimation of the BRS.

Let us explain the essence of the approach.

First we reconstruct variability by the variogram and its representation in frequency domain as previously explained by the (2.4) until to the (2.10). One time we will use this method on $RR$ time series and thus the same approach will be used for reconstructed $SBP$ time series. In this condition, we will introduce two new BRS indexes of variability, $BRS_{LF}$ and $BRS_{HF}$, for the two bands (LF) and (HF) respectively, calculated in the following manner:

$$BRS_{LF} = \sqrt{\frac{Total\ Variability\ calculated\ in\ the\ LF\ band\ for\ RR}{Total\ Variability\ calculated\ in\ the\ LF\ band\ for\ SBP}} \quad in\ \frac{m\sec.}{mmHg} \quad (3.1)$$

and

$$BRS_{HF} = \sqrt{\frac{Total\ Variability\ calculated\ in\ the\ HF\ band\ for\ RR}{Total\ Variability\ calculated\ in\ the\ HF\ band\ for\ SBP}} \quad in\ \frac{m\sec.}{mmHg} \quad (3.2)$$

A more accurate investigation, to be performed in a following paper, will require to use variogram values calculated by using the mean blood pressure time series given by $((SBP-DBP)/3+DBP)$ instead of SBP.

By using the CZF method we have also the possibility to investigate the coupling of the variabilities existing simultaneously, at each lag, between $RR$ and $SBP$. This important result may be praised, estimating also the corresponding coupling strength. The theory runs in the following manner. Given the two time series, one for $RR$ and the other for $SBP$, (we indicate them here by $x_i(t)$ and $y_i(t)$), the simultaneous coupling of their variabilities and the value of the coupling strength will be estimate in the following manner [Conte 2007] :

$$C_{RR-SBP}(h) = \frac{\gamma_{RR-SBP}(h)}{\sqrt{\gamma_{RR}(h)\gamma_{SBP}(h)}} \quad (3.3)$$

with $\gamma_{RR-SBP}(h)$ calculated as it follows:

$$\gamma_{RR-SBP}(h) = \frac{1}{2}E[[x_i(t+h)-x_i(t)][y_i(t+h)-y_i(t)]] \quad (3.4)$$

From the (3.3) it is seen that the maximum value for coupling strength is 0.5.

The set of the two indexes $BRS_{LF}$ and $BRS_{HF}$ plus the estimation of the coupling strength $C_{RR-SBP}(h)$, performed for each selected lag $h$, enables now us to perform a complete analysis of baroreflex sensitivity, based on the variability of the data, and calculated this time in a manner that may be retained to be free from approximations. In addition, the diagram $C_{RR-SBP}(h)$ vs the lag $h$ (or its representation in $Hz$ in the frequency domain) enables use to inspect the behaviour of the $RR-SBP$ coupling in time or in frequency domain with the further possibility to evaluate the sign of the simultaneous variability between $RR$ and $SBP$ by direct inspection of the regimes in which such variabilities will result both positive or both negative or positive and negative respectively: ($RR$ var.(+), $SBP$ var(+)), ($RR$ var.(−), $SBP$ var(−)), ($RR$ var.(+), $SBP$ var(−)), ($RR$ var.(−), $SBP$ var(+)). It is evident that each of the four mentioned regimes, delineate a definite physiological condition.

This last explanation completes the exposition of our method.

## 4. Methods

Seven healthy subjects (4 women, 3 men) (age included between 21 and 28 years old) underwent continuous noninvasive blood pressure (BP) recording using a Finapres 2300 device together with an ECG recording using a BioPac system. Subjects were recorded 10-12 min in standing position. In order to evaluate the great variability of the involved signals and of the estimated indexes, we studied each subject two times with a time interval of a week for a total of fourteen examined subjects.

Data were provided as BP and ECG signals sampled at 250 Hz with a 16-bit resolution. Systolic BP (SBP) values were obtained by software for peak maximum identification, and R-R values were obtained by software for time intervals identification in ECG. Pieces of 400 points were selected for analysis of variability. Lags in the variogram calculation were considered from 1 to 397.

## 5. Estimation of BRS

The BRS estimates included the procedure exposed in the previous section, obtained by the CZF method. To this method we added also the investigation performed by the traditional spectral analysis previously discussed in the framework of the standard methods used currently to calculate BRS. Still, it is well known that R-R time series represent data unevenly sampled. Traditional Fast Fourier Transform (FFT) requires instead that time series should be sampled at equally spaced time intervals. The Fourier spectrum of unevenly sampled data requires the use of Lomb-Scargle periodogram technique [Lomb 1976, Scargle 1989]. In order to give back a more accurate comparison of the BRS estimation by the CZF method and that one of Fourier spectral analysis, we performed also Lomb-Scargle analysis. Obviously, all the three methods provided to estimate BRS values in the low-frequency (LF) or high-frequency (HF) bands.

## 6. Results

The results of variability analysis based on CZF method are given in Figures 1,2,3 and in Tables 1, 2. Figure 1 shows variability of RR time intervals in unities (sec$^2$) for subject….. , and it is expressed in frequency domain in the range from zero to 0.4 Hz. Figure 2 shows variability of SBP in unities (mmHg$^2$) for the same subject in frequency domain and in the same range. Finally, Figure 3 evidences for the same subject the coupling strength connected to simultaneous variability of $RR$ and $SBP$ in the same frequency domain.

Table 1 gives instead the calculated values of $BRS_{LF}$ and $BRS_{HF}$, calculated respectively by the CZF method, by using the Fourier spectrum and by employing the Fourier spectrum for unevenly sampled data by the Lomb-Scargle technique. Finally, Table 2 gives the values of the coupling strength, taken in modulus, for simultaneous variability of $RR$ and $SBP$. Table 3 gives classification of the investigated subjects for age and sex.

It is seen that the present calculation based on analysis of variability of $RR$ and $SBP$ gives values for variability based BRS estimations that in some manner result in certain cases quite similar to the calculations obtained by the traditional spectral analysis performed by FFT and Lomb-Scargle while instead in other cases the two methods effort very different values. FFT and Lomb-Scargle give instead results that are often in quite satisfactory agreement. We retain that our calculations, based on CZF, are given free from drastic approximations as instead it results in the case of Fourier transformations and Lomb-Scargle. Therefore, we conclude that they represent a better representation of the physiological dynamics under consideration, and thus it should be currently adopted in such BRS investigations. The next step is now to apply the present methodology with particular application of the (2.8) and the (2.9) for fractal analysis in cases of clinical interest.

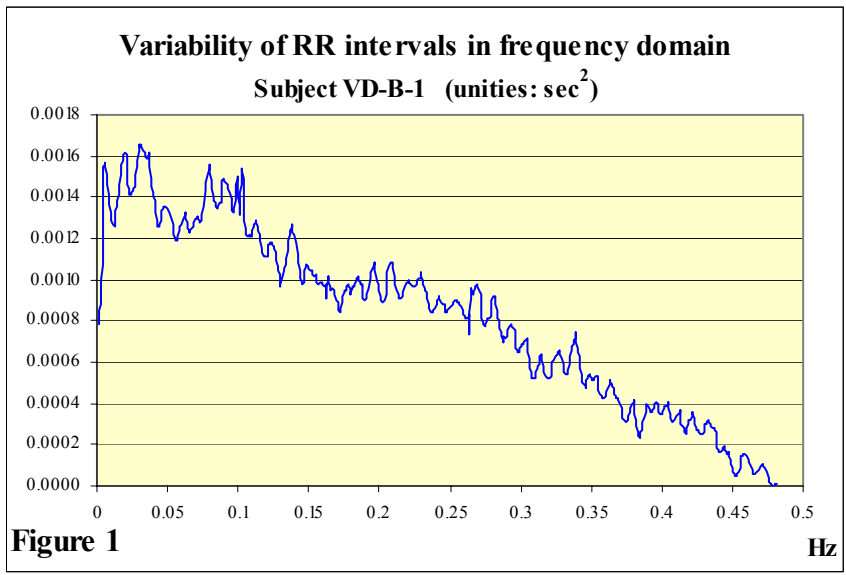

Figure 1

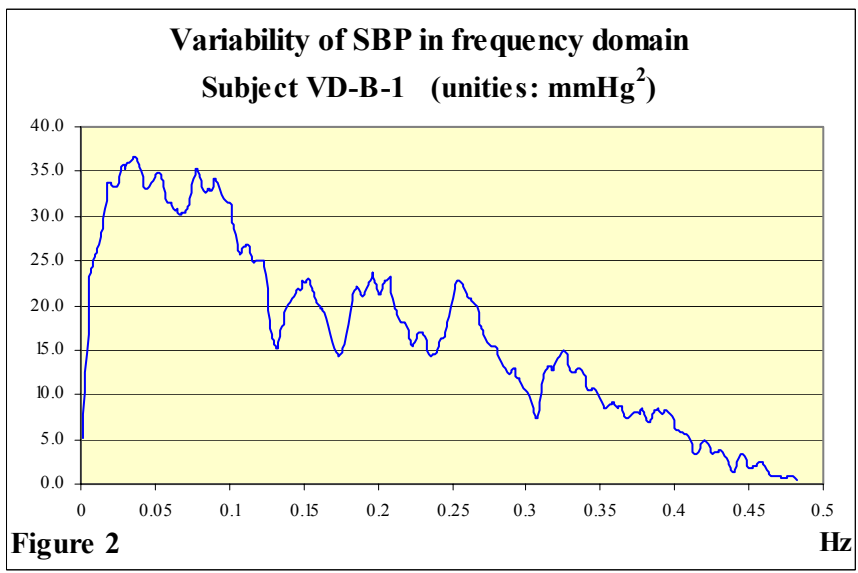

Figure 2

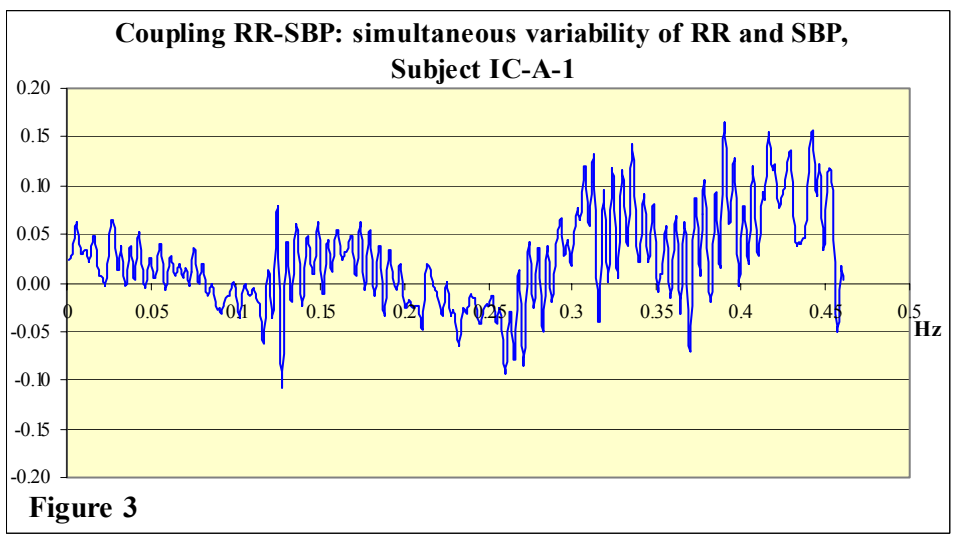

Figure 3

| Table 1: Estimation of BaroReflex Sensitivity (BRS) by different methods (unities: msec/mmHg) ||||||||
|---|---|---|---|---|---|---|---|
| Subject | | CZF Method || Fourier Spectral Method || Lomb-Scargle ||
| | | $BRS_{LF}$ | $BRS_{HF}$ | $BRS_{LF}$ | $BRS_{HF}$ | $BRS_{LF}$ | $BRS_{HF}$ |
| AM-A-1 | | 7.414 | 7.264 | 6.445 | 6.327 | 6.327 | 6.347 |
| CR-A-1 | | 4.523 | 3.632 | 10.475 | 4.147 | 10.834 | 4.266 |
| DPC-A-1 | | 7.012 | 5.367 | 8.636 | 8.896 | 8.848 | 9.425 |
| IC-A-1 | | 24.877 | 24.340 | 28.175 | 34.804 | 28.399 | 34.250 |
| RR-A-1 | | 24.661 | 27.874 | 28.169 | 31.060 | 28.385 | 30.253 |
| SC-A-1 | | 14.284 | 12.301 | 14.190 | 17.042 | 14.637 | 18.285 |
| VD-A-1 | | 3.758 | 4.233 | 5.737 | 4.497 | 5.758 | 4.529 |
| AM-B-1 | | 9.433 | 9.125 | 8.190 | 18.054 | 8.114 | 18.177 |
| CR-B-1 | | 8.885 | 8.247 | 10.935 | 21.551 | 11.637 | 21.601 |
| DPC-B-1 | | 5.456 | 4.831 | 15.003 | 21.934 | 9.392 | 9.321 |
| IC-B-1 | | 21.959 | 22.082 | 18.778 | 31.635 | 19.388 | 32.240 |
| RR-B-1 | | 26.350 | 20.140 | 33.454 | 36.362 | 33.822 | 36.537 |
| SC-B-1 | | 10.389 | 6.351 | 11.791 | 22.349 | 11.617 | 23.009 |
| VD-B-1 | | 6.688 | 7.097 | 8.016 | 10.967 | 8.117 | 11.068 |

Table 2: Coupling strength for simultaneous variability of *RR* and *SBP*

| Subject | $BRS_{LF}$ | $BRS_{HF}$ | $C^{RR-SBP}_{LF}$ | $C^{RR-SBP}_{HF}$ |
|---|---|---|---|---|
| AM-A-1 | 7.414 | 7.264 | 12.639 | 26.005 |
| CR-A-1 | 4.523 | 3.632 | 6.238 | 19.739 |
| DPC-A-1 | 7.012 | 5.367 | 14.396 | 32.535 |
| IC-A-1 | 24.877 | 24.340 | 2.041 | 8.601 |
| RR-A-1 | 24.661 | 27.874 | 4.669 | 12.797 |
| SC-A-1 | 14.284 | 12.301 | 4.879 | 26.386 |
| VD-A-1 | 3.758 | 4.233 | 11.182 | 34.457 |
| AM-B-1 | 9.433 | 9.125 | 5.589 | 17.958 |
| CR-B-1 | 8.885 | 8.247 | 5.841 | 19.559 |
| DPC-B-1 | 5.456 | 4.831 | 7.694 | 24.546 |
| IC-B-1 | 21.959 | 22.082 | 3.088 | 9.589 |
| RR-B-1 | 26.350 | 20.140 | 1.645 | 5.468 |
| SC-B-1 | 10.389 | 6.351 | 11.427 | 34.965 |
| VD-B-1 | 6.688 | 7.097 | 24.752 | 50.613 |

**Table 3: subjects data.**

| Subject | Sex | Age (years) |
|---------|-----|-------------|
| AM-A-1  | m   | 21          |
| CR-A-1  | f   | 28          |
| DPC-A-1 | m   | 21          |
| IC-A-1  | f   | 21          |
| RR-A-1  | f   | 22          |
| SC-A-1  | f   | 23          |
| VD-A-1  | m   | 23          |
| AM-B-1  | m   | 21          |
| CR-B-1  | f   | 28          |
| DPC-B-1 | m   | 21          |
| IC-B-1  | f   | 21          |
| RR-B-1  | f   | 22          |
| SC-B-1  | f   | 23          |
| VD-B- 1 | m   | 23          |